\documentclass[12pt]{article}
\usepackage{a4wide,epsfig,amsmath,amssymb,cite,scalefnt,graphicx}

\def\npb{{Nucl.\ Phys.\ }{\bf B}}
\def\be{\begin{equation}}
\def\ee{\end{equation}}
\def\bea{\begin{eqnarray}}
\def\eea{\end{eqnarray}}
\def\nn{\nonumber\\}

\def\tr{\hbox{tr}}

\def\prd{{Phys.\ Rev.\ }{\bf D}}

\def\prl{Phys.\ Rev.\ Lett.\ }

\def\thetabar{{\overline\theta}}
\def\alphadot{\dot\alpha}
\def\betadot{\dot\beta}
\def\Qbar{\overline Q}
\def\psibar{\overline{\psi}}

\def\frak#1#2{{\textstyle{{#1}\over{#2}}}}

\def\ybar{\overline y}

\def\frak#1#2{{\textstyle{{#1}\over{#2}}}}

\def\lambdabar{\overline\lambda}

\def\Dtil{\tilde D}
\def\Dhat{\hat D}
\def\Ftil{\tilde F}

\def\sigmabar{\overline\sigma}
\def\phibar{\overline\phi}

\def\psibar{\overline\psi}
\def\Fbar{\overline F}

\def\Dtil{\tilde D}

\def\Ncal{{\cal N}}
\def\Ftil{\tilde F}
\def\alphadot{\dot\alpha}
\def\betadot{\dot\beta}

\def\pa{\partial}

\input epsf
\begin{document}

\begin{titlepage}
\begin{flushright}
LTH838\\
\end{flushright}
\date{}
\vspace*{3mm}

\begin{center}
{\Huge
A renormalisable
non-anticommutative supersymmetric $SU(N)\otimes U(1)$ gauge theory in 
components}\\[12mm]{\bf I.~Jack, D.R.T.~Jones and R.~Purdy}\\

\vspace{5mm}
Dept. of Mathematical Sciences,
University of Liverpool, Liverpool L69 3BX, UK\\

\end{center}

\vspace{3mm}
\begin{abstract}
We discuss the non-anticommutative ($\Ncal=\frak12$)
supersymmetric $SU(N)\otimes U(1)$ gauge theory 
including a superpotential. We show how recent proposals
for obtaining a renormalisable version of the theory may be implemented
in the component formalism at the one-loop level.

\end{abstract}

\vfill

\end{titlepage}

\section{\label{intro}Introduction}
Deformed quantum field theories have been subject to
renewed attention in recent years due to their natural appearance
in string theory. Initial investigations focussed on theories on  
non-commutative spacetime in which the commutators of the spacetime
co-ordinates become
non-zero. More recently\cite{casal,casala,brink,schwarz,ferr,klemm,abbas,deboer,
oog}, non-anticommutative supersymmetric theories have been
constructed by deforming the anticommutators of the Grassmann co-ordinates
$\theta^{\alpha}$
(while leaving the anticommutators of the $\thetabar{}^{\alphadot}$ unaltered).
Consequently, the anticommutators of the supersymmetry generators
$\Qbar_{\alphadot}$ are
deformed while those of the $Q_{\alpha}$ are unchanged. It is straightforward
to construct non-anticommutative versions of ordinary supersymmetric theories
by taking the superspace action and replacing the ordinary product by
the Moyal $*$-product\cite{seib} which implements the non-anticommutativity.
Non-anticommutative versions of the Wess-Zumino model and supersymmetric gauge  
theories have been formulated in four
dimensions\cite{seib,araki} and their renormalisability
discussed\cite{brittoa,brittoaa,terash,brittob,rom,lunin},
with explicit computations up to two loops\cite{grisa} for the Wess-Zumino
model and one loop for gauge theories\cite{jjwa,jjwb,penrom,grisb,jjwc}.
Even more recently, non-anticommutative theories in two dimensions have been
constructed\cite{inami, chand, chanda, luis,chandc}, and
their one-loop divergences computed\cite{arakib,jp}.
In Ref.~\cite{jjp}
we returned to a closer examination of the non-anticommutative
Wess-Zumino model (with a superpotential) in four dimensions, and
showed that to obtain correct
results for the theory where the auxiliary fields have been eliminated,
from the corresponding results for
the uneliminated theory, it is necessary to include
in the classical action separate couplings for all the terms which may be
generated by the renormalisation process; and finally in Ref.~\cite{jjpa}
we extended this analysis to the gauged $U(1)$ case.

In Ref.~\cite{jjwc} we considered the renormalisation
of an $\Ncal=\frak12$ theory with a superpotential (for the case of adjoint 
matter) and with a mass term (for the case of matter in the fundamental
and anti-fundamental 
representations); note that $\Ncal=\frak12$ supersymmetry 
does not allow a trilinear term in the latter case. 
We found there were obstacles to obtaining
a renormalisable theory with a superpotential in the adjoint case.
The requirements of $\Ncal=\frak12$ invariance and renormalisability
impose the choice of gauge group $SU(N)\otimes U(1)$ (rather than $SU(N)$ or
$U(N)$)\cite{jjwa}, \cite{jjwb}. In the adjoint case with a
trilinear superpotential, the matter fields must also be 
in a representation of $SU(N)\otimes U(1)$. The problem is that the 
potential part of the classical 
action contains terms with different combinations of $SU(N)$ and $U(1)$ chiral
fields which mix under $\Ncal=\frak12$ supersymmetry, but for which the 
Yukawa couplings renormalise differently, at least in the simplest version 
of the theory. However, recently an elegant solution to this problem has been
proposed\cite{penroma} in which the kinetic terms for the $U(1)$ chiral fields 
are modified, in such a way that the $SU(N)$ and $U(1)$ chiral fields (and 
consequently their Yukawa couplings) renormalise in exactly the same way.
The authors of Ref.~\cite{penroma} worked in superspace; our purpose here is
to confirm that a similar procedure can be carried out in the component
formalism. 

\section{The classical adjoint action}
In this section we present the classical form of the
adjoint $\Ncal=\frak12$ action with a superpotential in the component 
formalism, including the modifications suggested in Ref.~\cite{penroma}. 
The adjoint action was first 
introduced in Ref.~\cite{araki}\ for the gauge group 
$U(N)$. However, 
as we noted in Refs.~\cite{jjwa}, \cite{jjwb}, 
at the quantum level the $U(N)$ gauge 
invariance cannot be retained since the $SU(N)$ and $U(1)$ gauge couplings 
renormalise differently; and we are
obliged to consider a modified $\Ncal=\frak12$ invariant theory  
with the gauge group $SU(N)\otimes U(1)$. In the adjoint case with a 
Yukawa superpotential, 
it turns out that the matter fields must also be in the adjoint
representation of $SU(N)\otimes U(1)$. 
The classical action with a superpotential may be written
\bea
S_0&=&\int d^4x
\Bigl\{e^{AB}(-\frak14F^{\mu\nu A}F^B_{\mu\nu}-i\lambdabar^A\sigmabar^{\mu}
(D_{\mu}\lambda)^B+\frak12D^AD^B)\nn
&&-\frak12iC^{\mu\nu}d^{ABC}e^{AD}F^D_{\mu\nu}\lambdabar^B\lambdabar^C\nn
&&+\Fbar F-i\psibar\sigmabar^{\mu}D_{\mu}\psi-D^{\mu}\phibar D_{\mu}\phi
+\phibar D^F\phi +
i\sqrt2(\phibar \lambda^F\psi-\psibar\lambdabar{}^F\phi)\nn
&&+C^{\mu\nu}(
\sqrt2D_{\mu}\phibar\lambdabar{}^D\sigmabar_{\nu}\psi
+i\phibar F^D_{\mu\nu} F)\nn
&&+
(\kappa-1)\bigl[
\Fbar^0 F^0
-i\psibar^0\sigmabar^{\mu}\pa_{\mu}\psi^0-\pa^{\mu}\phibar^0 \pa_{\mu}\phi^0\nn
&&+d^{000}C^{\mu\nu}(
\sqrt2\pa_{\mu}\phibar^0\lambdabar{}^0\sigmabar_{\nu}\psi^0
+i\phibar^0 F^0_{\mu\nu} F^0)\nn      
&&+d^{ab0}C^{\mu\nu}(\sqrt2D_{\mu}\phibar^a\lambdabar{}^b
\sigmabar_{\nu}\psi^0+i\phibar^a F^b_{\mu\nu} F^0)
\bigr]\nn
&&+\frak12\left(yd^{ABC}\phi^A\phi^B F^C
-yd^{ABC}\phi^A\psi^B\psi^C
+\ybar d^{ABC}\phibar^A\phibar^B \Fbar^C
-\ybar d^{ABC}\phibar^A\psibar^B\psibar^C\right)\nn
&&+\frak13i\ybar C^{\mu\nu}f^{abc}D_{\mu}\phibar^a D_{\nu}\phibar^b\phibar^c
-\frak{1}{3}i\ybar C^{\mu\nu}d^{ABE}d^{CDE}
F^D_{\mu\nu}\phibar^A\phibar^B\phibar^C\nn
&&+\kappa_1\sqrt2C^{\mu\nu}d^{abc}(\phibar^a\lambdabar^b\sigmabar_{\nu}
D_{\mu}\psi^c
+D_{\mu}\phibar^a\lambdabar^b\sigmabar_{\nu}\psi^c+i\phibar^aF_{\mu\nu}^b
F^c)\nn
&&+\kappa_2\sqrt2C^{\mu\nu}d^{ab0}(\phibar^0\lambdabar^a\sigmabar_{\nu}
D_{\mu}\psi^b
+\pa_{\mu}\phibar^0\lambdabar^a\sigmabar_{\nu}\psi^b
+i\phibar^0F_{\mu\nu}^aF^b)\nn
&&+\kappa_3\sqrt2C^{\mu\nu}d^{ab0}(\phibar^a\lambdabar^b\sigmabar_{\nu}
\pa_{\mu}\psi^0
+D_{\mu}\phibar^a\lambdabar^b\sigmabar_{\nu}\psi^0
+i\phibar^aF_{\mu\nu}^bF^0)\nn
&&+\kappa_4\sqrt2C^{\mu\nu}d^{0ab}(\phibar^a\lambdabar^0\sigmabar_{\nu}
D_{\mu}\psi^b
+D_{\mu}\phibar^a\lambdabar^0\sigmabar_{\nu}\psi^b
+i\phibar^aF_{\mu\nu}^0
F^b)\nn
&&+\kappa_5\sqrt2C^{\mu\nu}d^{000}(\phibar^0\lambdabar^0\sigmabar_{\nu}
\pa_{\mu}\psi^0
+\pa_{\mu}\phibar^0\lambdabar^0\sigmabar_{\nu}\psi^0
+i\phibar^0F_{\mu\nu}^0
F^0)\Bigr\}.
\label{Sadj}
\eea
where
\bea
\lambda^F&=&\lambda^A\Ftil^A,\quad (\Ftil^A)^{BC}=if^{BAC},\nn
\lambda^D&=&\lambda^A \Dtil^A,\quad (\Dtil^A)^{BC}=d^{ABC},
\label{lamdef}
\eea
(similarly for $D^F$, $F_{\mu\nu}^D$), and we have
\bea
D_{\mu}\phi&=&\pa_{\mu}\phi+iA^F_{\mu}\phi,\nn
F_{\mu\nu}^A&=&\pa_{\mu}A_{\nu}^A-\pa_{\nu}A_{\mu}^A-f^{ABC}
A_{\mu}^BA_{\nu}^C,
\label{Dmudef}
\eea
with similar definitions for $D_{\mu}\psi$, $D_{\mu}\lambda$.
If one decomposes $U(N)$ as
$SU(N)\otimes U(1)$ then our convention is that $\phi^a$ (for example) 
are the $SU(N)$
components and $\phi^0$ the $U(1)$ component.
Of course then $f^{ABC}=0$
unless all indices are $SU(N)$. 
We note that $d^{ab0}=\sqrt{\frak2N}\delta^{ab}$, $d^{000}=\sqrt{\frak2N}$.
(Useful identities for $U(N)$ are listed in the Appendix.)
We also have
\be
e^{ab}=\frac{1}{g^2},\quad e^{00}=\frac{1}{g_0^2}, \quad e^{0a}=e^{a0}=0.
\label{etensor}
\ee
Compared with our previous work such as Ref.~\cite{jjwc}, we have absorbed a 
factor of $g$ into our definitions of the fields in the gauge multiplet.
For simplicity of exposition
we shall omit (here and elsewhere)
terms which are $\Ncal=\frak12$ supersymmetric on
their own (such as terms involving only $\phibar$, $\lambdabar$ 
and/or $F$). Such terms are present
in the action as obtained by reduction of the superspace action to components,
and they are also generated by quantum corrections even if omitted from the
classical action; but they do not add to our understanding of the 
renormalisability of the theory, which is our main concern here. They were 
considered in full in Refs.~\cite{penroma}; and indeed we
included them ourselves in Refs.~\cite{jjwa}, \cite{jjwb}.
We have, however, taken the opportunity of including here some additional
sets of terms (those multiplied by $\kappa_{1-5}$)
which will be required for renormalisability of the theory. Each of these sets
of terms is separately $\Ncal=\frak12$ invariant. Note that for the chiral
field kinetic part of the action 
in Eq.~(\ref{Sadj}), $\Fbar F\equiv \Fbar^A F^A=\Fbar^a F^a+\Fbar^0 F^0$, 
etc; the $U(1)$ part $\Fbar^0F^0$ 
could have been combined with that in the $(\kappa-1)$ part of the action,
as could the kinetic terms with $\phi^0$ and $\psi^0$, with some attendant
simplification.
We have left the action in its present form to facilitate comparison with
Ref.~\cite{penroma}.

It is easy to show that Eq.~(\ref{Sadj})\ is invariant under
\bea
\delta A^A_{\mu}&=& -i\lambdabar^A\sigmabar_{\mu}\epsilon\nn  
\delta \lambda^A_{\alpha}&=& i\epsilon_{\alpha}D^A+\left(\sigma^{\mu\nu}\epsilon
\right)_{\alpha}\left[F^A_{\mu\nu}
+\frak12iC_{\mu\nu}d^{ABC}\lambdabar^B\lambdabar^C\right],\quad
\delta\lambdabar^A_{\alphadot}=0,\nn
\delta D^A&=& -\epsilon\sigma^{\mu}D_{\mu}\lambdabar^A,\nn
\delta\phi&=& \sqrt2\epsilon\psi,\quad\delta\phibar=0,\nn
\delta\psi^{\alpha}&=& \sqrt2\epsilon^{\alpha} F,\quad
\delta\psibar_{\alphadot}=-i\sqrt2(D_{\mu}\phibar)
(\epsilon\sigma^{\mu})_{\alphadot},\nn
\delta F^A&=& 0,\nn
\delta \Fbar^A&=& -i\sqrt2D_{\mu}\psibar^A\sigmabar^{\mu}\epsilon
-2i(\phibar\epsilon\lambda^F)^A
+2C^{\mu\nu}D_{\mu}(\phibar^B\epsilon\sigma_{\nu}   
(\lambdabar^D)^{AB}).
\label{newsusy}
\eea
In Eq.~(\ref{Sadj}), $C^{\mu\nu}$ is related to the non-anti-commutativity 
parameter $C^{\alpha\beta}$ by  
\be
C^{\mu\nu}=C^{\alpha\beta}\epsilon_{\beta\gamma}
\sigma^{\mu\nu}_{\alpha}{}^{\gamma},
\label{Cmunu}
\ee
where 
\bea
\sigma^{\mu\nu}&=&\frak14(\sigma^{\mu}\sigmabar^{\nu}-
\sigma^{\nu}\sigmabar^{\mu}),\nn
\sigmabar^{\mu\nu}&=&\frak14(\sigmabar^{\mu}\sigma^{\nu}-
\sigmabar^{\nu}\sigma^{\mu}).
\label{sigmunu}
\eea
Our conventions are in accord with \cite{seib}; in particular, 
\be
\sigma^{\mu}\sigmabar^{\nu}=-\eta^{\mu\nu}+2\sigma^{\mu\nu}.
\label{sigid}
\ee
Properties of $C$ which follow from
Eq.~(\ref{Cmunu})\ are  
\bea
C^{\alpha\beta}&&=\frak12\epsilon^{\alpha\gamma}
\left(\sigma^{\mu\nu}\right)_\gamma{}^{\beta}C_{\mu\nu},
\nn
C^{\mu\nu}\sigma_{\nu\alpha\betadot}&&=C_{\alpha}{}^{\gamma}
\sigma^{\mu}{}_{\gamma\betadot},\nn
C^{\mu\nu}\sigmabar_{\nu}^{\alphadot\beta}&&=-C^{\beta}{}_{\gamma}
\sigmabar^{\mu\alphadot\gamma}.
\label{cprop}
\eea 
We use the standard gauge-fixing term 
\be
S_{\rm{gf}}={1\over{2\alpha}}\int d^4x (\pa.A)^2
\label{gafix}
\ee 
with its associated
ghost terms.  The vector propagator is given by  
\be
\Delta^{AB}_{V\mu\nu}=-{1\over{p^2}}\left(\eta_{\mu\nu}
+(\alpha-1){p_{\mu}p_{\nu}\over{p^2}}\right)\left(e^{-1}\right)^{AB}.
\label{gprop}
\ee
The scalar propagator is 
\be
\Delta_{\phi}^{AB}=-\frac{1}{p^2}P^{AB}
\label{sprop}
\ee
where
\be
P^{ab}=\delta^{ab},\quad P^{00}=\frac{1}{\kappa}, \quad P^{0a}=P^{a0}=0,
\ee
the fermion propagator is  
\be
\Delta^{AB}_{\psi\alpha\alphadot}=
{p_\mu\sigma^{\mu}_{\alpha\alphadot}\over{p^2}}P^{AB},
\label{fprop}
\ee
where the momentum enters at the end of the propagator with the undotted 
index,
and the auxiliary propagator is
\be
\Delta_F^{AB}=P^{AB}.
\label{aprop}
\ee

\section{Renormalisation}
The bare action will
be given as usual by replacing fields and couplings by their bare versions,
shortly to be given more explicitly.
Note that in the $\Ncal=\frak12$ supersymmetric case, fields and their
conjugates may renormalise differently. 
We found in Refs.~\cite{jjwa}, \cite{jjwb} 
that non-linear renormalisations of $\lambda$
and $\Fbar$ were required; and in a subsequent
paper\cite{jjwd} we pointed out that non-linear
renormalisations of $F$, $\Fbar$ are required even in ordinary $\Ncal=1$
supersymmetric gauge theory when working in the uneliminated formalism.
The renormalisations of the remaining fields and couplings are linear as 
usual (except for $\kappa$, $\kappa_{1-5}$, see later) and given by
\bea 
\lambdabar^a_B=Z_{\lambda}^{\frak12}\lambdabar^a,
\quad
A^{a}_{\mu B}=Z_A^{\frak12}A^{a}_{\mu}, 
&\quad& \phi^a_B=Z_{\phi}^{\frak12}\phi^a,\quad
\psi^a_B=Z_{\psi}^{\frak12}\psi^a,\nn
\phibar^a_B=Z_{\phi}^{\frak12}\phibar^a,
\quad \psibar^a_B=Z_{\psi}^{\frak12}\psibar^a, &\quad&
g_B=Z_gg,\quad y_B=Z_yy,\nn  
C_B^{\mu\nu}=Z_CC^{\mu\nu}, \quad (\kappa-1)_B&=&Z_{\kappa}(\kappa-1),\quad 
\kappa_{1-5B}=Z_{1-5}.
\label{bare}
\eea
The corresponding $U(1)$ gauge multiplet fields 
$\lambdabar^0$ etc are unrenormalised;
so is $g_0$. The renormalisation constants for the $U(1)$ chiral fields 
will be denoted $Z_{\phi^0}$ etc and discussed later.
In Eq.~(\ref{bare}), $Z_{1-5}$ are divergent
contributions; in other words we have set the renormalised couplings
$\kappa_{1-5}$ to zero for simplicity.   
The anomalous dimensions $Z_{\lambda}$ etc, and the renormalisation 
constants for the couplings $g$, $y$,
$C$ and $(\kappa-1)$, start with tree-level values of 1. (The slightly
non-standard definition of $Z_{\kappa}$ is once again to make our results
correspond more closely with those of Ref.~\cite{penroma}.) 
The one-loop graphs contributing
to the ``standard'' terms in the Lagrangian (those without a
$C^{\mu\nu}$) are the same as in the $\Ncal=1$ case, though we must now take 
into account the $\kappa$ dependence of the propagators for the $U(1)$ chiral
fields, as seen in Eqs.~(\ref{sprop}), (\ref{fprop}) and (\ref{aprop}); 
however, the 
anomalous dimensions for the gauge-multiplet fields
and hence the gauge $\beta$-functions are the same as in the standard 
$\Ncal=1$ theory. 
Since our gauge-fixing term in Eq.~(\ref{gafix})\ does not preserve 
supersymmetry, the anomalous dimensions for $A^a_{\mu}$ and $\lambda^a$
are
different (and moreover gauge-parameter dependent), as are those for
$\phi^a$ and $\psi^a$. However, the 
gauge $\beta$-functions are of course gauge-independent. 
We therefore have, at one loop\cite{timj}: 
\bea
Z_{\lambda}&=&1-2g^2NL(3+\alpha),\nn
Z_A&=&1-g^2NL(3+\alpha),\nn
Z_D&=&1-6g^2NL,\nn
Z_g&=&1-2g^2NL,
\label{Zgg}
\eea
where (using dimensional regularisation with $d=4-\epsilon$)
$L={1\over{16\pi^2\epsilon}}$; the results appear different from those in 
Ref.~\cite{timj} 
and indeed our earlier paper Ref.~\cite{jjwc} 
due to our absorption of the factor of $g$ into the gauge multiplet fields.

The divergent contributions 
corresponding to (for instance) the scalar kinetic terms take the form
\bea
&&L\left(-\tr[\Dtil^AP\Dtil^BP]y\ybar
\pa^{\mu}\phibar^A\pa_{\mu}\phi^B
+2g^2(1-\alpha)\pa^{\mu}\phibar^a\pa_{\mu}\phi^a\right)\nn
&=&
L\left\{-y\ybar\left[N+\frac{4}{N\kappa}(1-\kappa)\right]
+2g^2(1-\alpha)\right\}\pa^{\mu}\phibar^a\pa_{\mu}\phi^a\nn
&&-2Ly\ybar\left[N+\frac{1}{N\kappa^2}(1-\kappa^2)\right]
\pa^{\mu}\phibar^0\pa_{\mu}\phi^0
\label{keq1}
\eea
and this must be cancelled by
\be
-\left[Z_{\phi}\pa^{\mu}\phibar^a\pa_{\mu}\phi^a
+Z_{\phi^0}\pa^{\mu}\phibar^0\pa_{\mu}\phi^0
+Z_{\kappa}(\kappa-1)Z_{\phi^0}\pa^{\mu}\phibar^0\pa_{\mu}\phi^0\right]. 
\label{keq2}
\ee
(Here and elsewhere, when we mention divergent contributions, we mean
divergent contributions to the effective action.)
We immediately find (using similar results for the fermion and auxiliary kinetic
terms)
\bea
Z_{\phi}&=&\left\{-y\ybar\left[N+\frac{4}{N\kappa}(1-\kappa)\right]
+2g^2(1-\alpha)N\right\}L,\nn
Z_{\psi}&=&\left\{-y\ybar\left[N+\frac{4}{N\kappa}(1-\kappa)\right]
-2g^2(1+\alpha)N\right\}L,\nn
Z_F&=&-y\ybar\left[N+\frac{4}{N\kappa}(1-\kappa)\right]L.
\label{Zchi}
\eea
The assignment of $Z_{\phi^0}$ (and $Z_{\psi^0}$, $Z_{F^0}$) requires
more care (and note we are still at liberty to choose $Z_{\kappa}$).
Consider the $yd^{abc}\phi^a\psi^b\psi^c$ term. The only diagrams
contributing to this are gauge dependent and give (as usual)
\be
-\frak12(7+3\alpha)LNg^2yd^{abc}\phi^a\psi^b\psi^c.
\ee
We then deduce that at one loop
\bea
Z^{(1)}_y&=&-\frac12Z^{(1)}_{\phi}-Z^{(1)}_{\psi}
-(7+3\alpha)LNg^2\nn
&=&-\frac32\left\{-y\ybar\left[N+\frac{4}{N\kappa}(1-\kappa)\right]
+4g^2N\right\}L=-\frac32Z_{\Phi}^{(1)},
\label{keq3}
\eea
where we recognise 
\be
Z_{\Phi}^{(1)}=\left\{-y\ybar\left[N+\frac{4}{N\kappa}(1-\kappa)\right]
+4g^2N\right\}L 
\label{zphi}
\ee
as the one-loop contribution to
the $SU(N)$ chiral superfield renormalisation constant. This is in accord
with the non-renormalisation theorem. (We should note that the discussion 
of renormalisation of the $F\phi^2$ and $\Fbar\phibar^2$
terms in the potential requires the 
non-linear renormalisations of $\Fbar$, $F$ which will be given explicitly
later.) 
In the usual ($\kappa=1)$ case, the Yukawa terms involving (for instance)
$\phi^0\psi^b\psi^c$ would renormalise differently from 
the $y\phi^a\psi^b\psi^c$ term due to the difference between 
$Z_{\phi}$ and $Z_{\phi^0}$, and the different diagrams contributing
to the two terms, and would need a different Yukawa coupling,
$y'$ say, for renormalisability.   
To be precise, we would have (in analogy with Eq.~(\ref{keq3}), and
again invoking the non-renormalisation theorem)
\be
Z^{(1)}_{y'}=-\frac12Z^{(1)}_{\Phi^0}-Z^{(1)}_{\Phi}.
\ee 
On the other hand, the $\Ncal=\frak12$
supersymmetry transformations mix these two groups of terms and require
them to have the same coupling. It therefore seems impossible to achieve 
simultaneously both renormalisability and $\Ncal=\frak12$ supersymmetry.
The ingenious solution suggested in Ref.~\cite{penroma}
is to exploit the presence of 
$\kappa$ to adjust $Z_{\Phi^0}$ to match $Z_{\Phi}$. This then guarantees
that $y$ and $y'$ may be identified.
Moreover we note that the
difference between $Z_{\Phi}$ and $Z_{\phi}$, $Z_{\psi}$ is due solely 
to the choice of a non-supersymmetric gauge; the gauge-independent terms are
the same, and since there are no gauge interactions for the $U(1)$ fields 
anyway, we have 
\be
Z_{\phi^0}=Z_{\psi^0}=Z_{\Phi^0}.
\ee
We then find from Eqs.~(\ref{keq1}), (\ref{keq2}), and (\ref{zphi})
\bea
Z^{(1)}_{\kappa}&=&-{4g^2N\kappa\over{\kappa-1}}+
{y\ybar N(\kappa-2)\over{\kappa-1}}
-{2y\ybar(2\kappa^2-\kappa-1)\over{N\kappa^2}}.
\label{zlam}
\eea

\begin{figure}[H]
\includegraphics{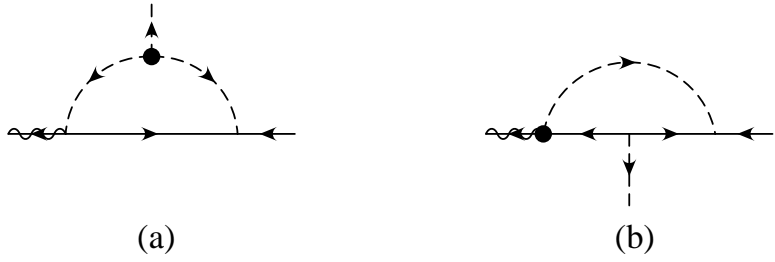}
\caption{\it Diagrams with one gaugino, one scalar and one
chiral fermion line (and two Yukawa couplings); 
the dot represents the position of a $C$.}\label{fig1} 
\end{figure}

\begin{figure}
\includegraphics{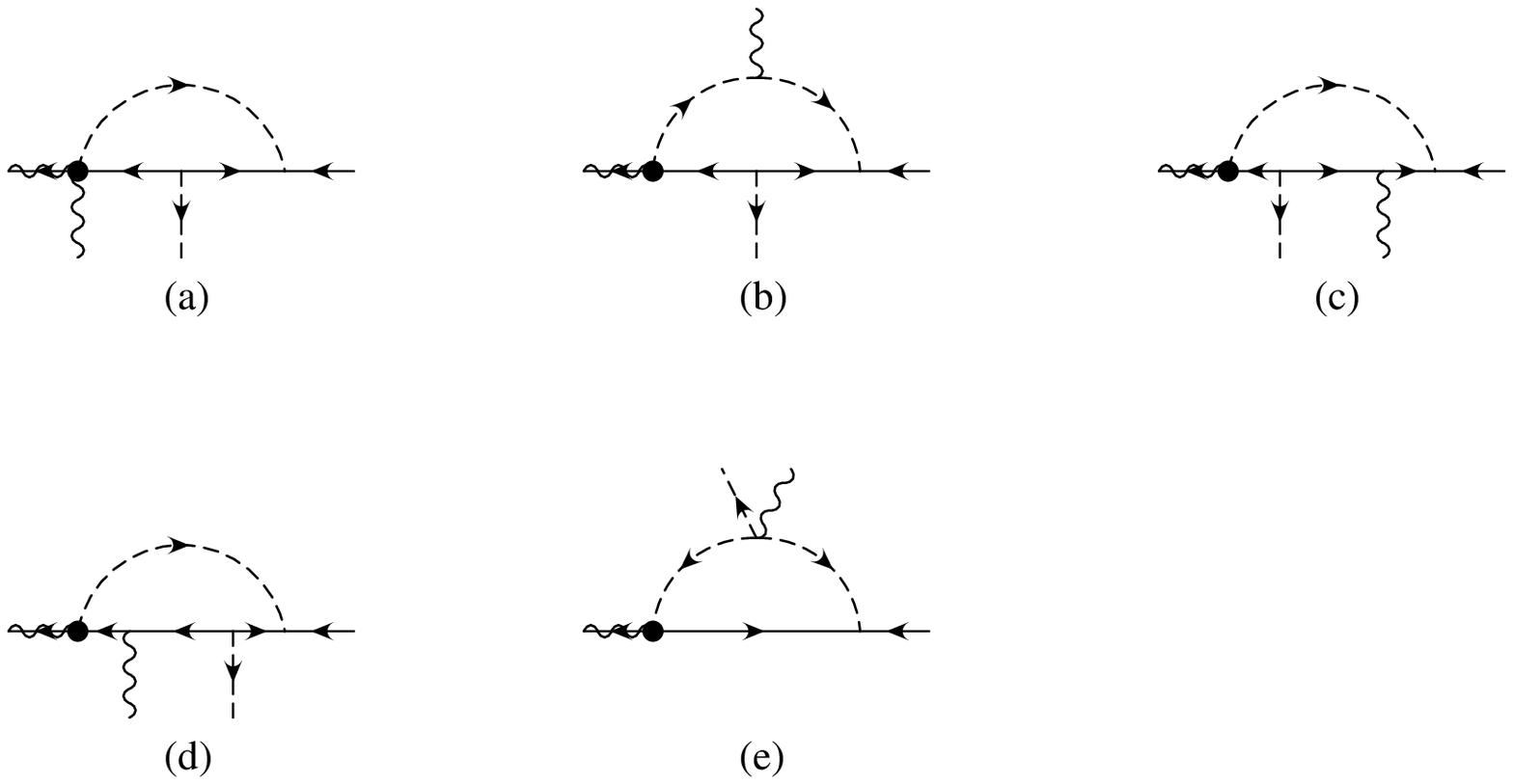}
\caption{\it Diagrams with one gaugino, one scalar, one
chiral fermion and one gauge line (and two Yukawa couplings).}
\label{fig2}
\end{figure}

\begin{figure}
\includegraphics{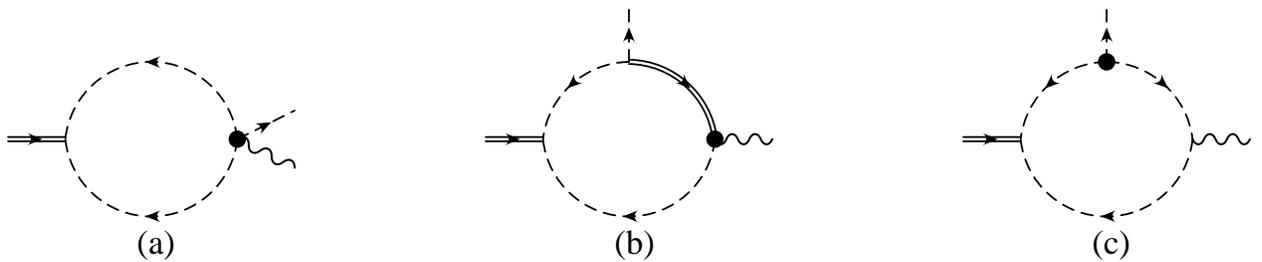}
\caption{\it Diagrams with one auxiliary, one scalar and one
gauge line (and two Yukawa couplings).}\label{fig3}
\end{figure}

\begin{figure}
\includegraphics{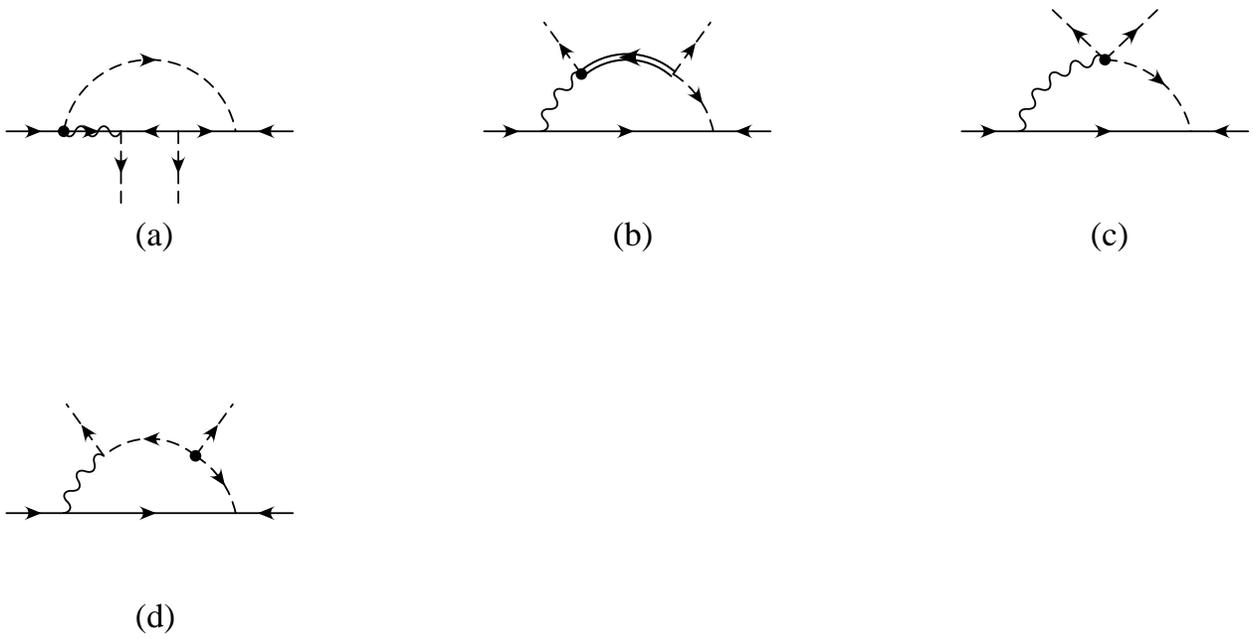}
\caption{\it Diagrams with two chiral fermion lines and two
scalars (and two Yukawa couplings).}\label{fig4}
\end{figure}

\begin{figure}
\includegraphics{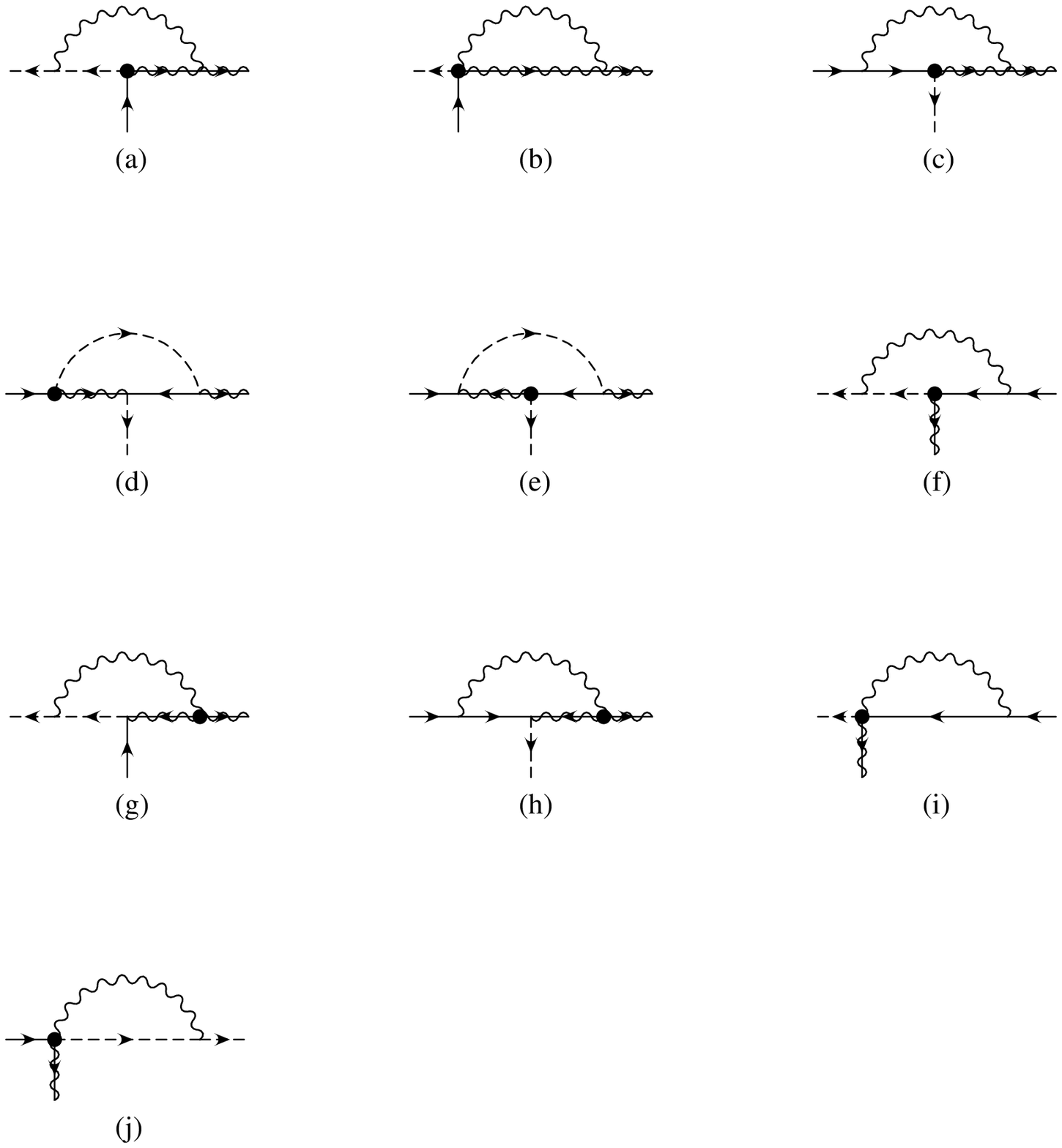}
\caption{\it Diagrams with one gaugino, one scalar and one
chiral fermion line.}\label{fig5}
\end{figure}

\begin{figure}
\includegraphics{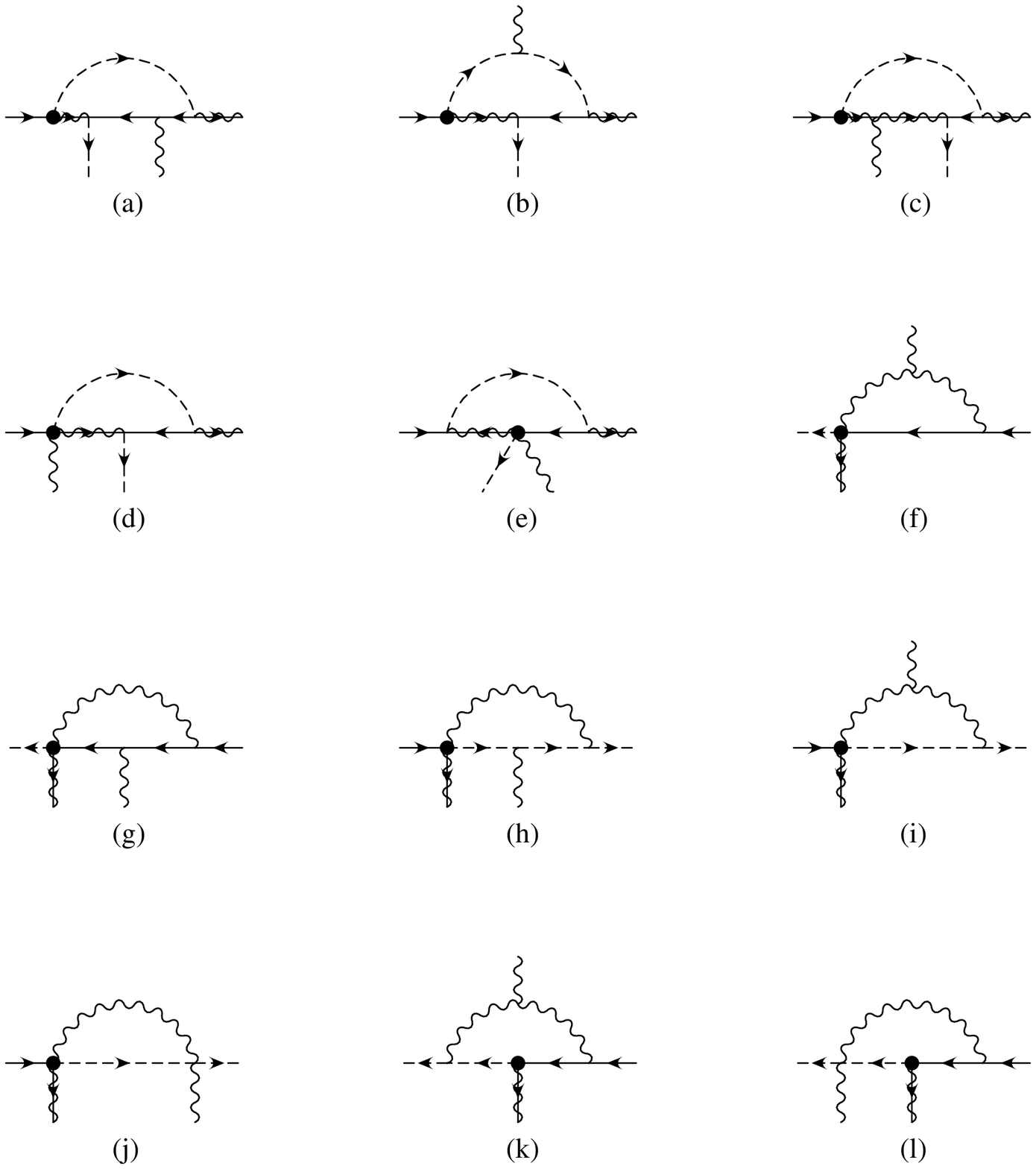}
\caption{\it Diagrams with one gaugino, one scalar, one
chiral fermion and one gauge line.}
\label{fig6}
\end{figure}

\begin{figure}
\includegraphics{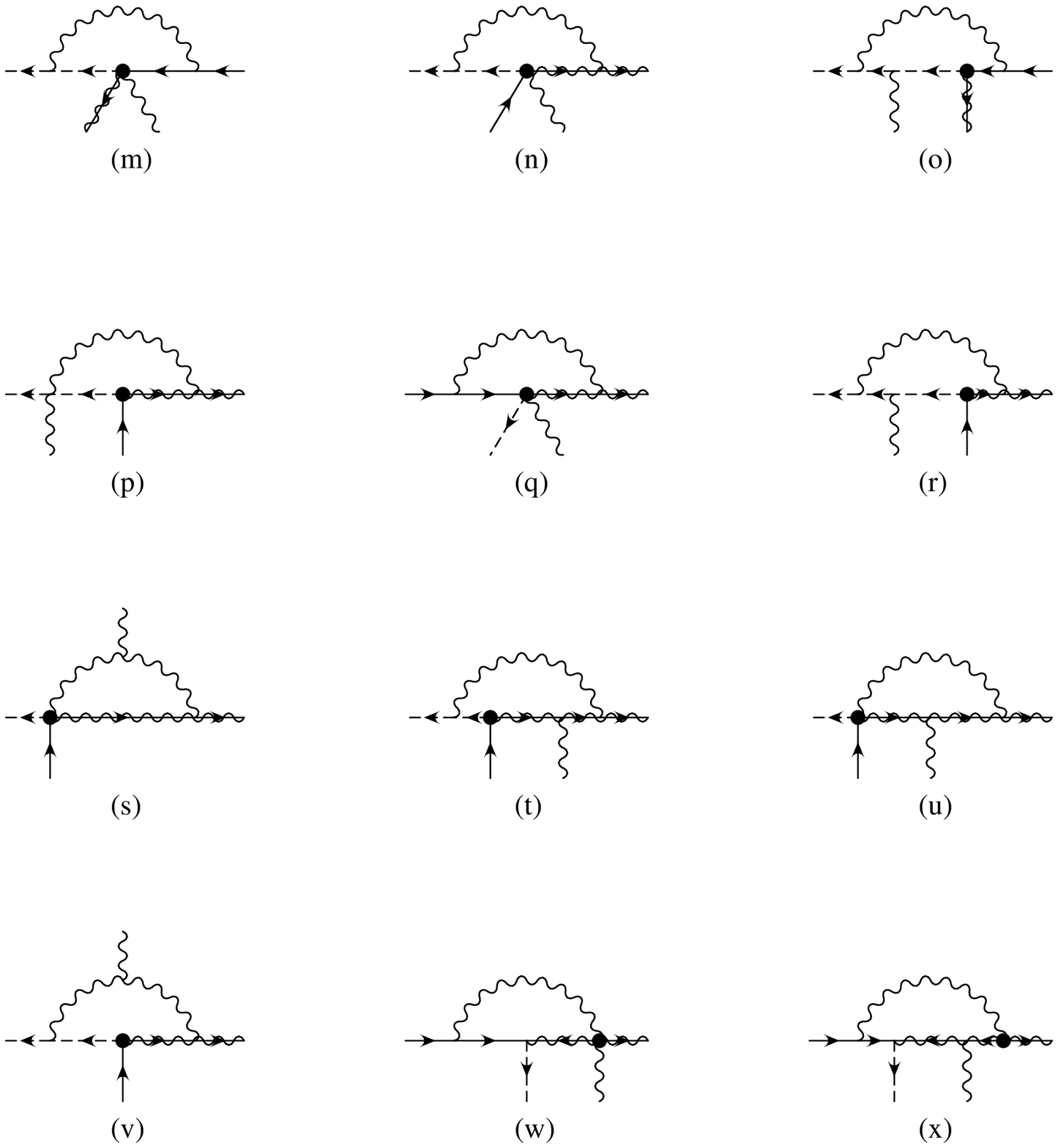}
\caption{\it Diagrams with one gaugino, one scalar, one
chiral fermion and one gauge line (continued).}
\label{fig7}   
\end{figure}

\begin{figure}
\includegraphics{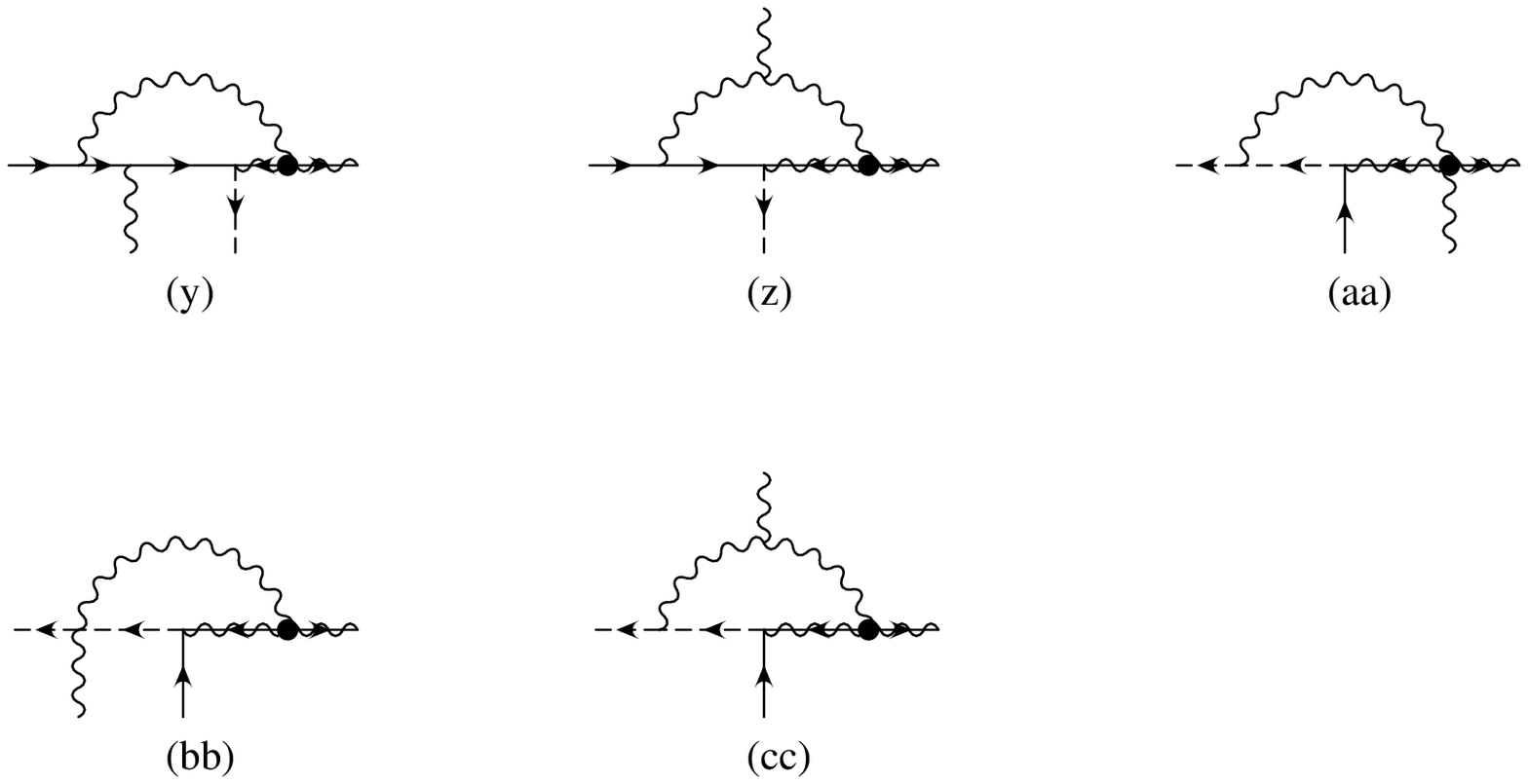}
\caption{\it Diagrams with one gaugino, one scalar, one
chiral fermion and one gauge line (continued).}
\label{fig8}   
\end{figure}

\begin{figure}
\includegraphics{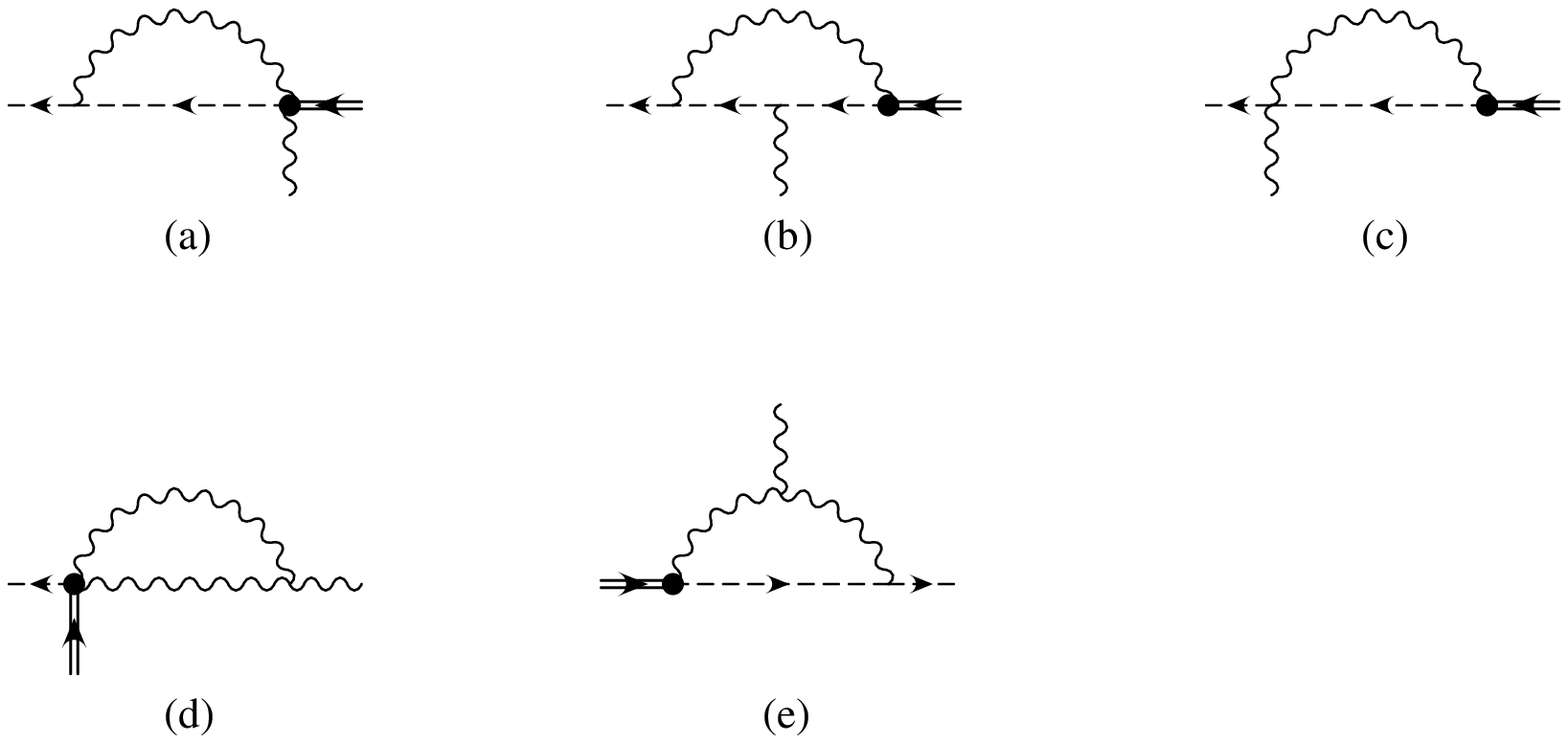}
\caption{\it Diagrams with one gauge, one scalar and one
auxiliary line.}
\label{fig9}
\end{figure}

\begin{figure}
\includegraphics{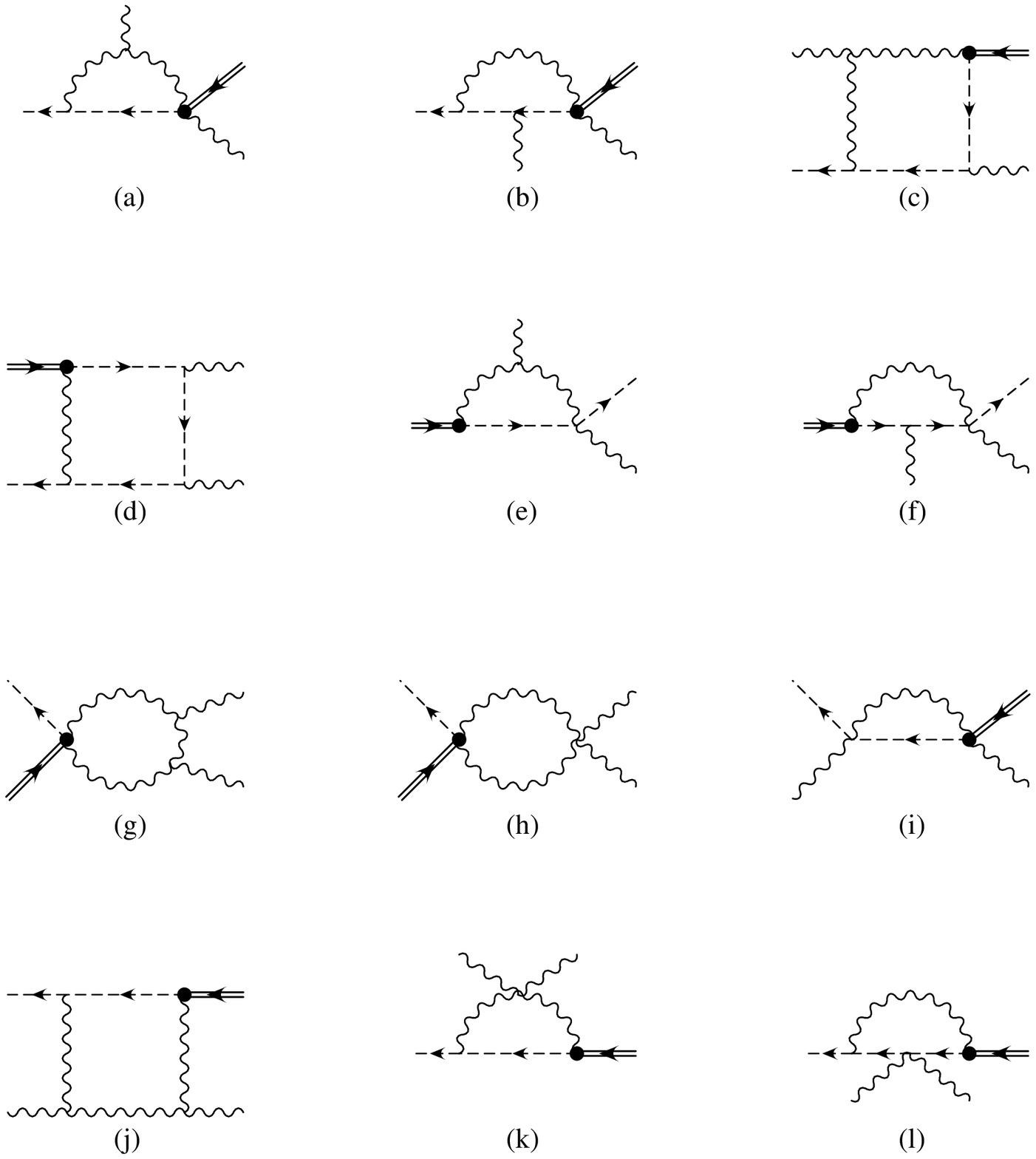}
\caption{\it Diagrams with two gauge, one scalar and one
auxiliary line.}
\label{fig10}
\end{figure}

\begin{figure}
\includegraphics{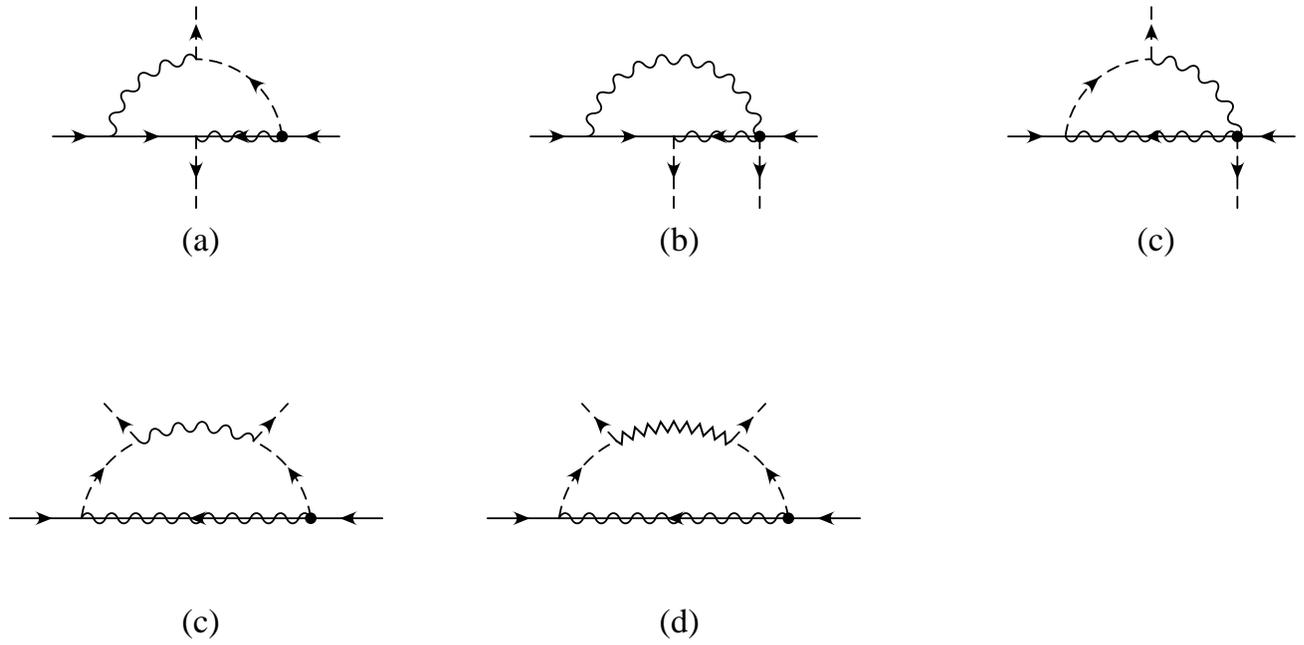}
\caption{\it \noindent Diagrams with two scalar
and two chiral fermion lines.}
\label{fig11}
\end{figure}

\begin{figure}
\includegraphics{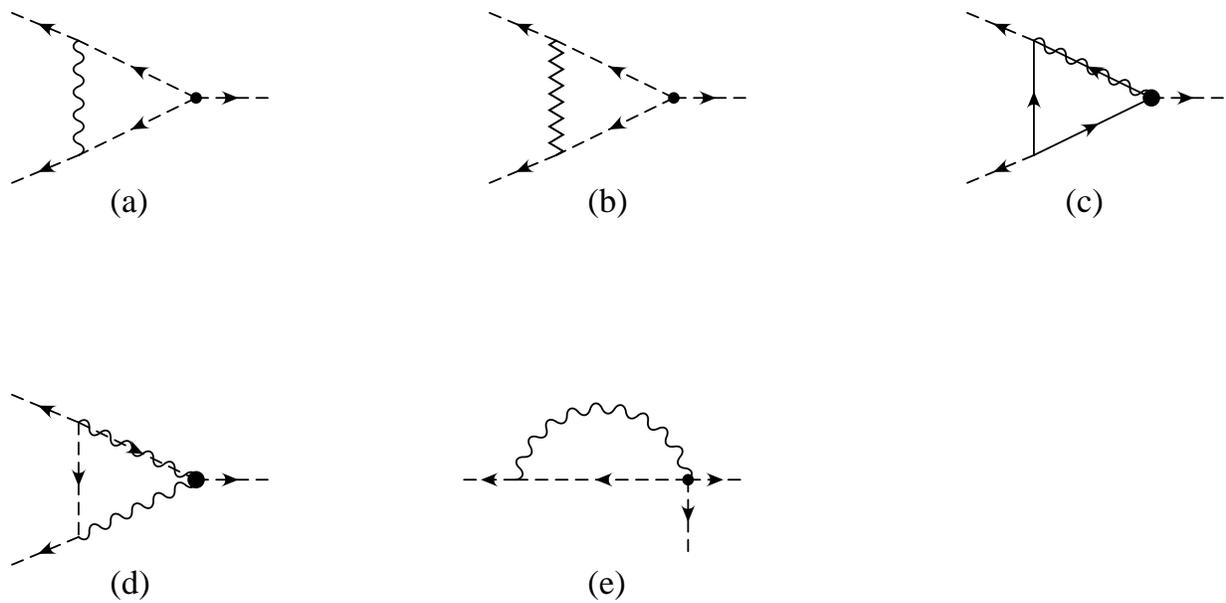}
\caption{\it Diagrams with three scalar lines.}
\label{fig12}
\end{figure}

\begin{figure}
\includegraphics{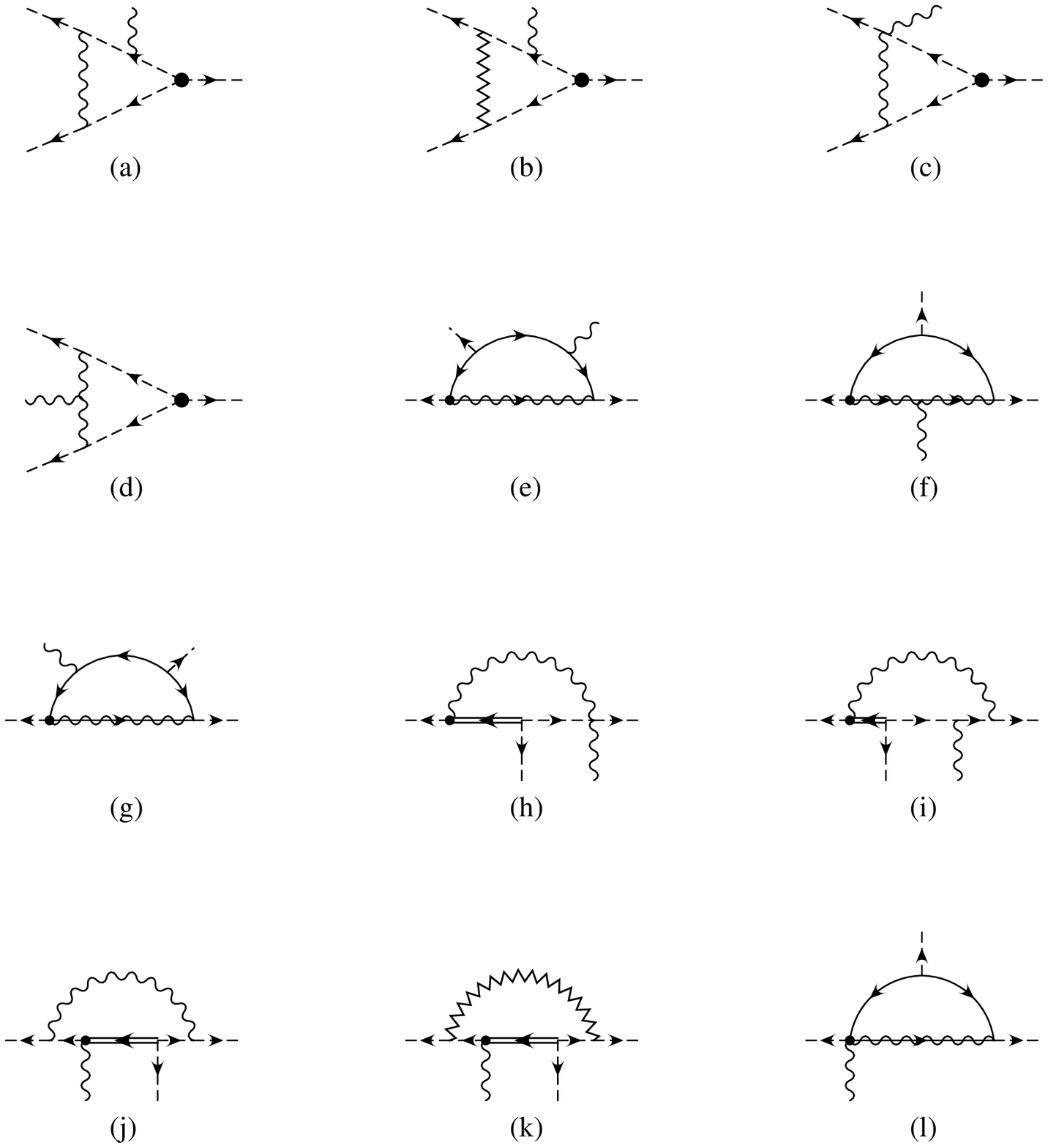}
\caption{\it Diagrams with three scalar,
one gauge line.}
\label{fig13}
\end{figure}

\begin{figure}
\includegraphics{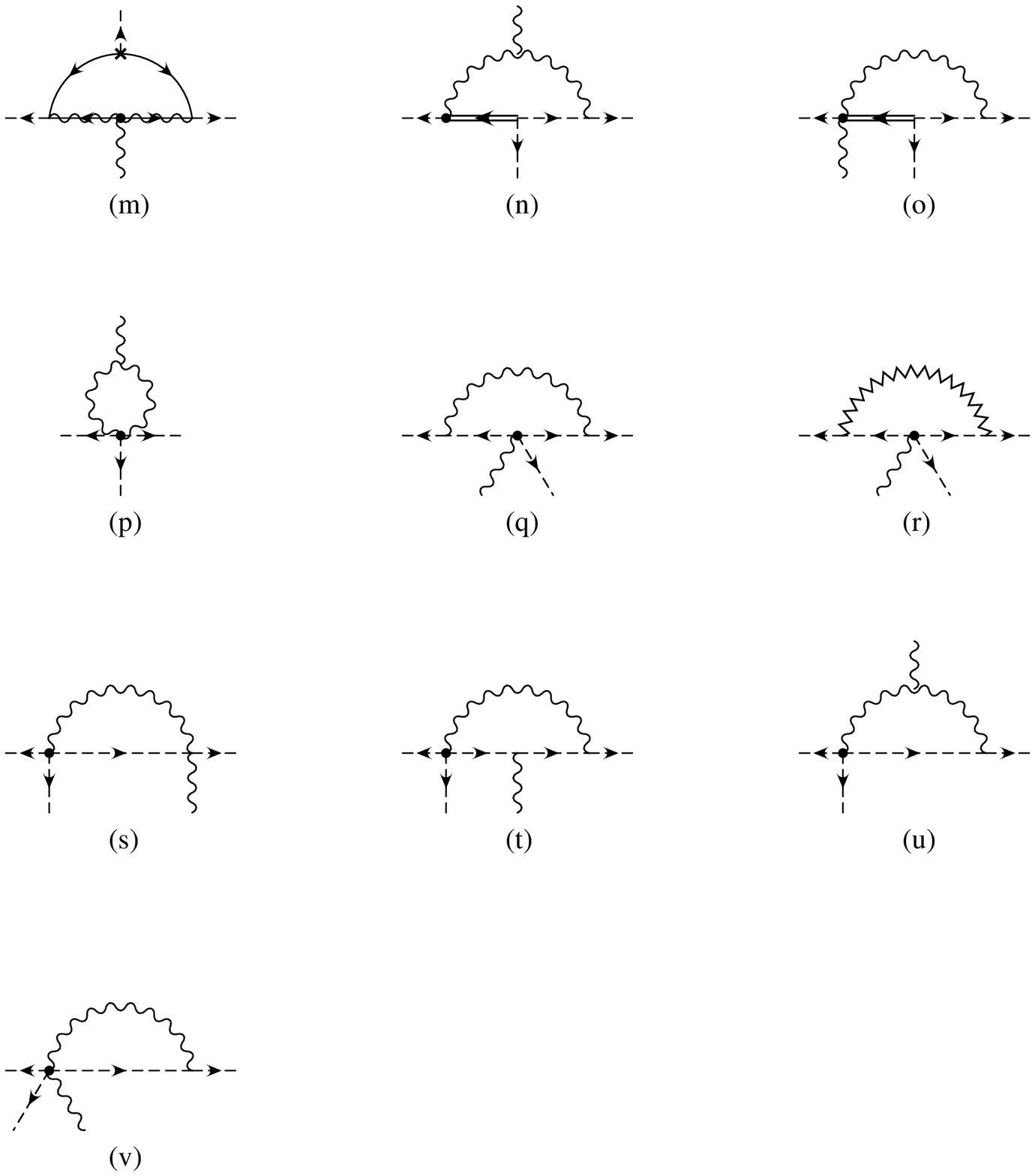}
\caption{\it Diagrams with three scalar,
one gauge line (continued)}   
\label{fig14}
\end{figure}

We have now dealt with the majority of the renormalisations of fields and 
couplings.  
The remaining non-linear renormalisations of $\lambda$, $F$ and $\Fbar$ are
largely determined in order to cancel $C$-dependent 
divergences; though as we have emphasised, 
a non-linear renormalisation of $F$ and $\Fbar$ is required
in the usual $\Ncal=1$ ($C=0$) case, and we shall quote the result of
Ref.~\cite{jjwd}.
So we now need to show how the $C$-dependent divergences are modified in the 
presence of $\kappa$ and check that we can choose these
non-linear renormalisations, together with $\kappa_{1-5}$, so that the theory
is renormalisable. In particular we shall verify that 
with our choice of $Z_{\kappa}$
and the identification of $Z_{\Phi^0}$ with $Z_{\Phi}$, the full 
potential (which includes $C$-dependent terms)
is indeed renormalisable with a single Yukawa coupling (though this is in 
principle guaranteed since the non-renormalisation theorem is known to
extend to the $\Ncal=\frac12$ case\cite{brittoaa}).  
The relevant divergent one-loop $C$-dependent graphs are depicted in 
Figs.~\ref{fig1}-\ref{fig14}. Figs.~\ref{fig1}-\ref{fig4} are graphs
giving contributions proportional to $y\ybar$. Figs.~\ref{fig1}-\ref{fig3}
were not computed by us previously in the adjoint case; we did 
compute Fig.~\ref{fig4}, but in any case the result needs reassessing in the 
present case with our $\kappa$-dependent action,
and will be radically different. Hence we shall shortly give a 
complete tabulation of the results for Figs.~\ref{fig1}-\ref{fig4}.
Figs.~\ref{fig5}--\ref{fig14} were all computed previously and in fact
we can obtain the results for our current $\kappa$-dependent action with
very simple modifications. We shall therefore simply present the results.

The divergent contributions from Fig.~\ref{fig1} are of the form 
\be
\sqrt2 C^{\mu\nu}y\ybar L
(X^{ABC}\pa_{\mu}\phibar^A\lambdabar^B\sigmabar_{\nu}\psi^C
+X^{\prime ABC}\phibar^A\lambdabar^B\sigmabar_{\nu}\pa_{\mu}\psi^C)
\ee
where $X^{ABC}$, $X^{\prime ABC}$ are as given in Table \ref{taba}.
\begin{table}
\begin{center}
\begin{tabular}{|c| c c|} \hline
&$X$&$X'$\\ \hline
a&$\tr[\Ftil^A\Ftil^B\Dtil^C]$&$0$ \\ \hline
b&$0$&$-\tr[\Dhat^BP\Dtil^AP\Dtil^CP]$ \\ \hline
\end{tabular}
\caption{\label{taba} Divergent contributions from Fig.~\ref{fig1}}
\end{center}
\end{table}
 
Here, 
\bea
(\Dhat^a)^{0b}=
(\Dhat^a)^{b0}=\kappa d^{ab0}&\quad& (\Dhat^0)^{00}=\kappa d^{000},\nn
(\Dhat^A)^{BC}&=&d^{ABC}\quad\hbox{otherwise}.
\label{Ddef}
\eea
Note that, although $P$ derives from the chiral field 
propagators in Eqs.~(\ref{sprop}), (\ref{fprop}), (\ref{aprop}), 
it is redundant when there is an $F$ on either side. 

The divergent contributions from Fig.~\ref{fig2} are of the form 
\be
\sqrt2 C^{\mu\nu}y\ybar L
Y^{ABCD}A^A_{\mu}\phibar^C\lambdabar^B\sigmabar_{\nu}\psi^D
\ee
where $Y^{ABCD}$ is as given in Table \ref{tabb}. 
\begin{table}
\begin{center}
\begin{tabular}{|c| c |} \hline
&$Y$\\ \hline
a&$2i\tr[\Ftil^A\Dhat^BP\Dtil^CP\Dtil^D]$ \\ \hline
b&$-i\tr[\Ftil^A\Dhat^BP\Dtil^CP\Dtil^D]$ \\ \hline
c&$-i\tr[\Ftil^A\Dtil^DP\Dhat^BP\Dtil^C]$ \\ \hline
d&$0$\\ \hline
e&$f^{ACX}\tr[\Ftil^X\Ftil^B\Dtil^D]$\\\hline
\end{tabular}
\caption{\label{tabb} Divergent contributions from Fig.~\ref{fig2}}
\end{center}
\end{table}

The contributions from Figs.~\ref{fig1}, \ref{fig2} add to
\bea
\Gamma_{1,2}^{(1)\rm{pole}}&=&y\ybar LC^{\mu\nu}\Bigl\{
-\frac12\left[N+\frac{8}{N\kappa}(1-\kappa)\right]d^{abc}
\phibar^a\lambdabar^b\sigmabar_{\nu}D_{\mu}\psi^c
+\frac{N}{2}D_{\mu}\phibar^a\lambdabar^b\sigmabar_{\nu}\psi^c\nn
&&-\left[N+\frac{4}{N\kappa}(1-\kappa)\right]d^{ab0}
(\phibar^a\lambdabar^0\sigmabar_{\nu}D_{\mu}\psi^b
+\phibar^0\lambdabar^a\sigmabar_{\nu}D_{\mu}\psi^b)\nn
&&-\left[N+\frac{2}{N\kappa^2}(1-\kappa^2)\right]d^{ab0}
\phibar^a\lambdabar^b\sigmabar_{\nu}\pa_{\mu}\psi^0
+Nd^{ab0}D_{\mu}\phibar^a\lambdabar^b\sigmabar_{\nu}\psi^0,\nn
&&-2\left[N+\frac{1}{N\kappa^2}(1-\kappa^2)\right]d^{000}    
\phibar^0\lambdabar^0\sigmabar_{\nu}\pa_{\mu}\psi^0\Bigr\}
\label{fig12res}
\eea

The divergent contributions from Fig.~\ref{fig3} are of the form 
\be
iC^{\mu\nu}y\ybar L(Z^{ABC}\pa_{\mu}\phibar^AA_{\nu}^BF^C
+Z^{\prime ABC}\phibar^A\pa_{\mu}A_{\nu}^BF^C)
\ee
where $Z^{ABC}$, $Z^{\prime ABC}$ are as given in Table \ref{tabc}. 
\begin{table}
\begin{center}
\begin{tabular}{|c| c c |} \hline
&$Z$&$Z'$\\ \hline
a&$2\tr[\Ftil^A\Ftil^B\Dtil^C]$&$\frak23\tr[\Ftil^A\Ftil^B\Dtil^C]
+\frak83\tr[\Dtil^CP\Dtil^B\Dtil^AP]$\\ 
&&$+\frak43\tr[\Dtil^CP\Dtil^DP]d^{DAB}$\\ \hline
b&$0$&$-4\tr[\Dtil^CP\Dhat^BP\Dtil^AP]$\\ \hline
c&$-2\tr[\Ftil^A\Ftil^B\Dtil^C]$ &$0$\\ \hline
\end{tabular}
\caption{\label{tabc} Divergent contributions from Fig.~\ref{fig3}}
\end{center}
\end{table}
They add to
\be
\Gamma_{3}^{(1)\rm{pole}}=
iy\ybar LC^{\mu\nu}\Bigl[\frac{N}{2}d^{abc}\phibar^aF_{\mu\nu}^bF^c
+Nd^{ab0}\phibar^aF_{\mu\nu}^bF^0\Bigr],
\ee
where we have assumed that the $\phibar A A F$ diagrams which 
we have not computed yield the gauge completion of the $\phibar (\pa A) F$
terms.
The contributions from Fig.~\ref{fig4} are given by
\be
y\ybar g^2LZ_1^{ABCD}(C\psi)^B\psi^A\phibar^C\phibar^D
\ee
where $Z_1^{ABCD}$ is as given in Table \ref{tabd}. 
\begin{table}
\begin{center}
\begin{tabular}{|c| c |} \hline
&$Z_1$\\ \hline
a&$-\tr[\Ftil^C\Dtil^BP\Dtil^AP\Dtil^D]$ \\ \hline
b&$\tr[\Ftil^B\Dhat^CP\Dtil^D\Dtil^A]$ \\ \hline
c&$-\frac13\left(\tr[\Ftil^B\Dtil^EP\Dtil^A]d^{CDE}+2\tr[\Ftil^B\Dtil^CP\Dtil^DP\Dtil^A]
-\tr[\Ftil^B\Ftil^C\Ftil^D\Dtil^A]\right)$ \\ \hline
d&$0$ \\ \hline
\end{tabular}
\caption{\label{tabd} Divergent contributions from Fig.~\ref{fig4}}
\end{center}
\end{table}
They add to 
\be
\Gamma_{4}^{(1)\rm{pole}}=
\left[N+\frac{4}{N\kappa}(1-\kappa)\right]y\ybar g^2L
\left[f^{abe}d^{cde}(C\psi)^b\psi^a\phibar^c\phibar^d
+2f^{abe}d^{0ce}(C\psi)^b\psi^a\phibar^c\phibar^0\right].
\ee

The contributions from Figs.~\ref{fig5}--\ref{fig14} are listed below.

\bea
\Gamma_{5}^{(1)\rm{pole}}&=&
Ng^2\sqrt2LC^{\mu\nu}\Bigl[(2+3\alpha)d^{abc}\pa_{\mu}\phibar^a 
\lambdabar^b\sigmabar_{\nu}\psi^c
-d^{abc}\phibar^a \lambdabar^b \sigmabar_{\nu}\pa_{\mu}\psi^c  \nn
&& +2\kappa(1+\alpha)d^{ab0}\pa_{\mu}\phibar^a 
\lambdabar^b\sigmabar_{\nu}\psi^0
-2\kappa d^{ab0}\phibar^a \lambdabar^b \sigmabar_{\nu}\pa_{\mu}\psi^0\nn
&& +2\alpha d^{ab0}\pa_{\mu}\phibar^a \lambdabar^0\sigmabar_{\nu}\psi^b\nn
&&+2(1+\alpha)d^{ab0}\pa_{\mu}\phibar^0 \lambdabar^a\sigmabar_{\nu}\psi^b
\Bigr] ,\nn
\Gamma_{6,7,8}^{(1)\rm{pole}}&=&\sqrt2
g^2LC^{\mu\nu}A_{\mu}^a\Bigl[
\left(\frak72(1+\alpha)f^{bae}d^{cde}-f^{dae}d^{cbe}+\frak12f^{bde}
d^{cae}\right)N
\phibar^b\lambdabar^c \sigmabar_{\nu}\psi^d\nn
&&-\frak12(1+5\alpha)\sqrt{2N}
f^{abc}\phibar^b\lambdabar^0 \sigmabar_{\nu}\psi^c
-\frak12\kappa
(7+5\alpha)\sqrt{2N}f^{abc}\phibar^b\lambdabar^c \sigmabar_{\nu}\psi^0
\Bigr],\nn
\Gamma_{9}^{(1)\rm{pole}}&=&
iNg^2LC^{\mu\nu}\Bigl[-(4-\alpha)d^{abc}
\phibar^b \pa_{\mu}A^a_{\nu}F^c\nn
&&-3\kappa(1-\alpha)d^{ab0}\phibar^a\pa_{\mu}A^b_{\nu}
F^0-
(5+\alpha)d^{ab0}\phibar^0\pa_{\mu}A^a_{\nu}F^b\Bigr],\nn
\Gamma_{10}^{(1)\rm{pole}}&=&ig^2
LC^{\mu\nu}A^a_{\mu}A^b_{\nu}
\Bigl(\frak14(3-4\alpha)Nf^{abe}d^{cde}\phibar^cF^d\nn 
&&-2\alpha\kappa\sqrt{2N}f^{abc}\phibar^cF^0   
+\frak32\sqrt{2N}f^{abc}\phibar^0F^c\Bigr),\nn
\Gamma_{11}^{(1)\rm{pole}}
&=&-iLNg^2C^{\alpha\beta}d^{abe}f^{cde}\phibar^a
\phibar^b\psi^c_{\alpha}\psi^d_{\beta},\nn
\Gamma_{12}^{(1)\rm{pole}}&=&
\frak12 N\ybar g^2LC^{\mu\nu}(1+\alpha)
f^{abc}\pa_{\mu}\phibar^a\pa_{\nu}\phibar^b\phibar^c
,\nn
\Gamma_{13,14}^{(1)\rm{pole}}&=&iC^{\mu\nu}\ybar g^2L\Bigl(
-\frak12\left(3+\frak73\alpha\right)Nf^{abe}f^{cde}\pa_{\mu}\phibar^a
\phibar^b\phibar^cA_{\nu}^d\nn
&&+\left[-\left(\frak54-\frak16\alpha\right)Nd^{abe}d^{cde}
+\left(3+\frak73\alpha\right)\delta^{ab}\delta^{cd}\right]
\phibar^a\phibar^b\phibar^c\pa_{\mu}A_{\nu}^d\nn
&&-\frak12(9+\alpha)\sqrt{2N}d^{abc}\phibar^0\phibar^a\phibar^b
\pa_{\mu}A_{\nu}^c-(5+\alpha)\phibar^0\phibar^0\phibar^a
\pa_{\mu}A_{\nu}^a\nn
&&-2\sqrt{2N}d^{abc}\phibar^a\phibar^b\phibar^c\pa_{\mu}A_{\nu}^0
-8\phibar^a\phibar^a\phibar^0\pa_{\mu}A_{\nu}^0\Bigr).
\eea

We now need to specify the remaining  renormalisations, 
of $F$, $\Fbar$ and $\lambda$, required to cancel the divergences.
The renormalisation of $\lambda^A$ is given by
\bea
\lambda_B^a&=&Z_{\lambda}^{\frak12}\lambda^a
-\frak12NLg^2C^{\mu\nu}d^{abc}
\sigma_{\mu}\lambdabar^cA_{\nu}^b
-NLg^2C^{\mu\nu}d^{ab0}
\sigma_{\mu}\lambdabar^0A_{\nu}^b\nn
&&+i\sqrt2\tau_1NLg^4d^{abc}(C\psi)^b\phibar^c
+i\sqrt2\tau_2NLg^4d^{ab0}(C\psi)^0\phibar^b,\nn
\lambda_B^0&=&\lambda^0+i\sqrt2\tau_3NLg^2g_0^2d^{0ab}
(C\psi)^a\phibar^b,
\label{lchange}
\eea
where $(C\psi)^{\alpha}=C^{\alpha}{}_{\beta}\psi^{\beta}$. 
The coefficients of the non-linear terms on the first line of 
Eq.~(\ref{lchange}) were computed in Ref.~\cite{jjwb});
the values of $\tau_{1-3}$
will be specified later. The replacement
of $\lambda$ by $\lambda_B$ produces a change in the action given (to
first order) by
\bea
S_0(\lambda_B)-S_0(\lambda)&=&NLg^2\int d^4x\Bigl\{
-\frak12f^{bde}d^{cae}A^a_{\mu}\phibar^b\lambdabar^c\sigmabar_{\nu}\psi^d
-f^{abe}d^{ec0}A^a_{\mu}\phibar^b\lambdabar^0\sigmabar_{\nu}\psi^c\nn
&&+\tau_1\bigl[ig^2d^{abe}f^{cde}\phibar^a\phibar^b\psi^c(C\psi^d)\nn
&&+\sqrt2C^{\mu\nu}d^{abc}\phibar^a\lambdabar^b\sigmabar_{\nu}D_{\mu}\psi^c
+\sqrt2C^{\mu\nu}d^{abc}D_{\mu}\phibar^a\lambdabar^b\sigmabar_{\nu}
\psi^c\bigr]\nn
&&+\tau_2\sqrt2C^{\mu\nu}d^{ab0}(\phibar^a\lambdabar^b\sigmabar_{\nu}
\pa_{\mu}\psi^0
+D_{\mu}\phibar^a\lambdabar^b\sigmabar_{\nu}\psi^0)\nn
&&+\tau_3\sqrt2C^{\mu\nu}d^{0ab}(\phibar^a\lambdabar^0\sigmabar_{\nu}
D_{\mu}\psi^b  
+D_{\mu}\phibar^a\lambdabar^0\sigmabar_{\nu}\psi^b)+\ldots\Bigr\},
\label{Schlamb}
\eea
where the ellipsis indicates terms depending solely on gauge or gaugino 
fields (which were given previously in Ref.~\cite{jjwb}). 

We now find that to render finite the contributions linear in $F$,
we also require
\begin{subequations}
\bea
\Fbar^a_B&=&Z_F\Fbar^a+
iC^{\mu\nu}Lg^2\Bigl\{N\Bigl[(5+2\alpha)
\pa_{\mu}A_{\nu}^b-\frak14(11+4\alpha)
f^{bde}A_{\mu}^dA_{\nu}^e\Bigr]\phibar^c d^{abc}\nn
&&+\sqrt{2N}\Bigl[2(2+\alpha)\pa_{\mu}A_{\nu}^a
-\frak12(5+2\alpha)
f^{abc}A_{\mu}^bA_{\nu}^c\Bigr]\phibar^0\nn
&&+2\sqrt{2N}(3+\alpha)
\pa_{\mu}A_{\nu}^0\phibar^a\Bigr\}
+(\alpha+3)g^2NL\frak14y d^{abc}\phi^b\phi^c\nn
&&+\frak12(\alpha+3)y g^2NLd^{ab0}\phi^b\phi^0
+\tau_4g^2yLf^{abc}(C\psi)^b\psi^c+\ldots,
\label{fbarredefA}\\
\Fbar^0_B&=&Z_F\Fbar^0\label{fbarredefB}\\
F^{a}_B&=&Z_FF^a+(\alpha+3)g^2NL\frak14\ybar d^{abc}\phibar^b\phibar^c
+\frak12(\alpha+3)\ybar g^2NLd^{ab0}\phibar^b\phibar^0+\ldots,
\label{fbarredefC}\\
F^{0}_B&=&Z_FF^0, \label{fbarredefD}
\eea
\end{subequations}
where the ellipsis stands for $\phibar\lambdabar\lambdabar$ terms which 
only affect the separately $\Ncal=\frak12$ independent terms which we are 
omitting anyway. We should mention 
here that in Eq.~(5.5) of
Ref.~\cite{jjwc} the $y\phi\phi$ and $\ybar\phibar\phibar$ 
terms in Eq.~(\ref{fbarredefA}),
 (\ref{fbarredefC}) were inadvertently interchanged.
Writing $Z_C^{(n)}$ for the $n$-loop 
contribution to $Z_C$, and so on, we set
\be
Z_n^{(1)}=z_nL.
\ee
We now find that with
\bea
z_C=0, \quad \tau_1&=&1,\quad \tau_2=-2, \quad \tau_3=4,\nn
z_1&=&\frac12\left[N+\frac{8}{N\kappa}(1-\kappa)\right]y\ybar,\nn
z_2&=&\left[N+\frac{4}{N\kappa}(1-\kappa)\right]y\ybar,\nn
z_3&=&\left[N+\frac{2}{N\kappa^2}(1-\kappa^2)\right]y\ybar+4g^2,\nn
z_4&=&\left[N+\frac{4}{N\kappa}(1-\kappa)\right]y\ybar-4g^2,\nn
z_5&=&2\left[N+\frac{1}{N\kappa^2}(1-\kappa^2)\right]y\ybar,\nn
\tau_4&=&\left[N+\frac{2}{N\kappa^2}(1-\kappa^2)\right].
\eea
the one-loop effective action is finite.
In particular, the same coupling $y$ is sufficient for the renormalisation
of the full set of potential terms; and also the same non-anticommutativity 
parameter $C^{\mu\nu}$ is sufficient throughout and remains unrenormalised
at one loop. This is in contrast to the situation in Ref.~\cite{jjwc}, where we
were obliged to introduce several different Yukawa couplings and also
different $C^{\mu\nu}$ parameters for different groups of terms.

We note that the groups of terms involving $\kappa_{1-5}$ have an analogue in 
Ref.~\cite{penroma}, in the groups of
terms involving (in their notation) $t_{1-5}$, each group again being separately
invariant. Explicit one-loop results are not given for $t_{1-5}$; in any case,
we should probably not expect precise agreement due to our different gauge 
choices. While on the topic of comparison of the component and superfield 
approaches, we should mention the calculation of Ref.~\cite{penromb}. There 
a three-field $U(1)$ $\Ncal=\frak12$ model is considered in the superfield 
context. However, there the 
chiral fields are in the adjoint representation, whereas in Ref.~\cite{jjpa} we 
considered a three-field $U(1)$ $\Ncal=\frak12$ model with the chiral fields 
having charges $q$, $-q$, $0$. At least as far as the non-gauge parts of 
the results are concerned, we appear to have agreement.

\section{Conclusions} 
We have confirmed by a component calculation the conclusion reached in 
Ref.~\cite{penroma}, namely that the general 
$SU(N)\otimes U(1)$ $\Ncal=\frak12$ theory with 
a superpotential may be rendered renormalisable by a judicious choice of 
kinetic term for the $U(1)$ fields such that the renormalisations of
the $U(1)$ and $SU(N)$ chiral superfields are equal, which ensures that a single
Yukawa coupling is sufficient. This solves the difficulties which we 
encountered in Ref.~\cite{jjwc}; apart from restoring renormalisability, 
we also are no longer obliged to introduce several different 
non-anticommutativity tensors $C^{\mu\nu}$, some of which require a non-zero
renormalisation. $C^{\mu\nu}$ is unrenormalised at one loop. Our component
calculation is perhaps technically simpler than the superfield one (though of 
course the brevity of the current paper owes much to our exploitation of 
previous results in Ref.~\cite{jjwc}, and the fact that we have not computed 
divergences corresponding to separately $\Ncal=\frak12$ invariant terms). 
However, this is offset by the 
awkwardness of the various non-linear renormalisations which are required. 
We should mention that we have checked that the computation can also be 
carried out in the eliminated formalism, i.e. after eliminating $F$, $\Fbar$
using their equations of motion.

Since $C^{\mu\nu}$ is now confirmed to be completely unrenormalised at one loop,
it seems to us that the most pressing direction for further investigation is to
see whether this property extends to two loops. However, $C^{\mu\nu}$ being
a self-dual tensor, problems concerned with extending the definition
of the alternating tensor $\epsilon^{\mu\nu\rho\sigma}$ away from four 
dimensions seem likely to arise when using dimensional regularisation beyond 
one loop. A promising alternative could be the use of differential 
regularisation\cite{freed}.

\section{Appendix}
Identities for $SU(N)$ useful for simplifying the divergent contributions
listed in the Tables are\cite{azcar}
\bea
\tr[\Dtil^a\Dtil^b]={N^2-4\over{N}}\delta^{ab},&\quad&
\tr[\Dtil^a\Dtil^b\Dtil^c]=\frac{N^2-12}{2N}d^{abc},\nn
\tr[\Ftil^a\Ftil^b\Dtil^c]=\frac{N}{2}d^{abc},&\quad&
\tr[\Ftil^a\Dtil^b\Dtil^c]=i{N^2-4\over{2N}}f^{abc},\nn
\tr[\Ftil^a\Ftil^b\Ftil^c\Dtil^d]&=&
i\frac{N}{4}(d^{abx}f^{cdx}+d^{cdx}f^{abx}),\nn
\tr[\Ftil^a\Dtil^b\Dtil^c\Dtil^d]&=&
\frac{N^2-12}{4N}f^{abx}d^{cdx}+\frac{N}{4}d^{abx}f^{cdx}\nn
&&+\frac{1}{N}(f^{adx}d^{cbx}-f^{acx}d^{bdx}).
\eea
\vfill
\eject

\end{document}